\documentclass[3p,times,twocolumn,fleqn]{elsarticle}
 \biboptions{comma,sort&compress}
 
\usepackage{graphicx}
\usepackage{here}
\usepackage{ecrc}

\usepackage{slashed}
\usepackage{bm}
\usepackage{float}

\volume{00}

\firstpage{1}

\journalname{Nuclear and Particle Physics Proceedings}

\runauth{Marco Frasca}

\jid{nppp}

\jnltitlelogo{Nuclear and Particle Physics Proceedings}

\usepackage{amssymb}
\usepackage[figuresright]{rotating}

\begin{document}

\begin{frontmatter}

%%
%%%%%%%%%%%%%%%%%%%%%%%%%%%%%%%%%%%%%%%%%%%%%%%%%
\title{ Confinement studies in QCD with Dyson-Schwinger equations }
 % \corref{cor0}}
 \cortext[cor0]{Talk given at 23th International Conference in Quantum Chromodynamics (QCD 20,  35th anniversary),  27 - 30 October 2020, Montpellier - FR}
 \author[label1,label2]{Marco Frasca\fnref{fn1}}
 %\fntext[fn1]{Speaker, Corresponding author.}
%  \cortext[cor0]{FAPESP CNPq-Brasil PhD student fellow.}
\ead{marcofrasca@mclink.it}
\address[label1]{Via Erasmo Gattamelata, 3, 
00176 Rome (Italy)}
%\address[label2]{Laboratoire
%Particules et Univers de Montpellier, CNRS-IN2P3, 
%Case 070, Place Eug\`
%Bataillon, 34095 - Montpellier, France.}
% \author[label3]{F. Fanomezana\corref{cor1}}
%  \cortext[cor1]{PhD student.}
%\ead{fanfenos@yahoo.fr}
%\address[label3]{Institute of High-Energy Physics of Madagascar (iHEP-MAD), University of Antananarivo, 
%Madagascar}
% \author[label2,label4]{S. Narison\fnref{fn1}}
%   \fntext[fn1]{Speaker, Corresponding author.}
%    \ead{snarison@yahoo.fr}

%\address[label4]{Madagascar consultant of the Abdus Salam International Centre for Theoretical Physics (ICTP), via Beirut 6,34014 Trieste, Italy .}
% \author[label3]{A. Rabemananjara\corref{cor1}}
%  \cortext[cor2]{Ph.D. student}
%\ead{achris\_01@yahoo.fr}

\pagestyle{myheadings}
\markright{ }
\begin{abstract}
We provide a study of quantum chromodynamics with the technique of Dyson-Schwinger equations in differential form. In this way, we are able to approach the non-perturbative limit and recover, with some approximations, the 't Hooft limit of the theory. Quark mass in the propagator term goes off-shell at low-energies signaling confinement. A condition for such occurrence in the theory is provided. 
\end{abstract}
% \begin{document}
\begin{keyword}  
%% keywords here, in the form: keyword \sep keyword

%% MSC codes here, in the form: \MSC code \sep code
%% or \MSC[2008] code \sep code (2000 is the default)

\end{keyword}

\end{frontmatter}
%%%%%%%%%%%%
%\vspace*{-1.5cm}
\section{Introduction}

Yang-Mills set of Dyson-Schwinger equations can be solved through a class of exact solutions of the 1-point function, similarly to the $\phi^4$ theory \cite{Frasca:2015yva}. Indeed, both theories can map each other. Adding quarks to the Lagrangian makes the theory no more amenable to an exact treatment. But, notwithstanding such a difficulty, full QCD can be treated with the identical approach through Dyson-Schwinger equations. 

The technique we use is due to Bender, Milton and Savage \cite{Bender:1999ek}. This method has the important advantage that the differential form of the Dyson-Schwinger equations is retained. This is especially useful when, as in our case, we have exact solutions for the 1-point and 2-point equations in the classical case.

Therefore, starting from the results for Yang-Mills theory without quarks, the set of Dyson-Schwinger equations for quantum chromodynamics (QCD) becomes amenable to a perturbative treatment in the strong coupling limit provided the 't Hooft limit $N\rightarrow\infty,\ Ng^2=constant,\ Ng^2\gg 1$ is taken.

Our main conclusion is that the low-energy limit of QCD yields a confining non-local Nambu-Jona-Lasinio approximation \cite{Frasca:2019ysi} when the condition for confinement is assumed to have the quark propagator off-shell due to the behaviour of the mass function of quarks \cite{Roberts:1994dr,Gribov:1998kb}. Therefore, no free quark is observable being not anymore a state of the theory. In this paper we will follow the derivation given in Ref.\cite{Frasca:2019ysi}.

\section{Bender-Milton-Savage method}

The principal point in the Bender-Milton-Savage (BMS) technique is to derive the Dyson-Schwinger equations retaining their PDE form \cite{Bender:1999ek}. In this way, vertexes are never introduced and there is no need to move to momenta space obtaining cumbersome integral expressions.

We consider the partition function of a given theory, e..g a scalar field theory to fix the ideas, given by
\begin{equation}
    Z[j]=\int[D\phi]e^{iS(\phi)+i\int d^4xj(x)\phi(x)}.
\end{equation}
For the 1P-function, it is
\begin{equation}
\left\langle\frac{\delta S}{\delta\phi(x)}\right\rangle=j(x)
\end{equation}
being
\begin{equation}
\left\langle\ldots\right\rangle=\frac{\int[D\phi]\ldots e^{iS(\phi)+i\int d^4xj(x)\phi(x)}}{\int[D\phi]e^{iS(\phi)+i\int d^4xj(x)\phi(x)}}
\end{equation}
Then, we set $j=0$. We derive this equation again with respect to $j$ to get the equation for the 2P-function. We assume the following definition of the nP-functions
\begin{equation}
\langle\phi(x_1)\phi(x_2)\ldots\phi(x_n)\rangle=\frac{\delta^n\ln(Z[j])}{\delta j(x_1)\delta j(x_2)\ldots\delta j(x_n)}.
\end{equation}
This will yield
\begin{equation}
\frac{\delta G_k(\ldots)}{\delta j(x)}=G_{k+1}(\ldots,x).
\end{equation}

This procedure can be iterated to any desired order giving, in principle, all the hierarchy of the Dyson-Schwinger equations in PDE form \cite{Frasca:2015yva}. This is advantageous when the solutions for 1P- and 2P-functions are known in the classical case.

\section{1P and 2P functions of QCD}

For our computations, we choose the Landau gauge that permits to simplify the computations and decouples the ghost field.

By applying the BMS tecnhinque to the QCD partition function \cite{Frasca:2019ysi}, one has for the 1P-functions
\begin{eqnarray}
      &&\partial^2G_{1\nu}^{a}(x)+gf^{abc}(
		\partial^\mu G_{2\mu\nu}^{bc}(0)+ \nonumber \\
		&&\partial^\mu G_{1\mu}^{b}(x)G_{1\nu}^{c}(x)-
		\partial_\nu G_{2\mu}^{\nu bc}(0)
		\nonumber \\
		&&-\partial_\nu G_{1\mu}^{b}(x)G_{1}^{\mu c}(x)) \nonumber \\
		&&+gf^{abc}\partial^\mu G_{2\mu\nu}^{bc}(0)+gf^{abc}\partial^\mu(G_{1\mu}^{b}(x)G_{1\nu}^{c}(x))
		\nonumber \\
		&&+g^2f^{abc}f^{cde}(G_{3\mu\nu}^{\mu bde}(0,0)
		+G_{2\mu\nu}^{bd}(0)G_{1}^{\mu e}(x)
		\nonumber \\
	    &&+G_{2\nu\rho}^{eb}(0)G_{1}^{\rho d}(x)
	    +G_{2\mu\nu}^{de}(0)G_{1}^{\mu b}(x)+ \nonumber \\
	    &&G_{1}^{\mu b}(x)G_{1\mu}^{d}(x)G_{1\nu}^{e}(x)) \nonumber \\
		&&=g\sum_{q,i}\gamma_\nu T^aS_{q}^{ii}(0)+g\sum_{q,i}{\bar q}_1^i(x)\gamma_\nu T^a q_1^i(x),
\end{eqnarray}
and for the quarks
\begin{equation}
	(i\slashed\partial-{\hat M}_q)q_{1}^{i}(x)+g{\bm T}\cdot\slashed{\bm G}_1(x) q_{1}^{i}(x) = 0,
\end{equation}
the mass function being given by
\begin{equation}
    {\hat M}_q^i=m_qI-g{\bm T}\cdot\slashed{\bm W}^{i}_q(x,x).
\end{equation}
Here and in the following Greek indexes ($\mu,\nu,\ldots$) are for the space-time and Latin index ($a, b,\ldots$) for the gauge group.
It is not difficult to see that, as expected, in the equations for the correlation functions of lower order appear contributions from higher order correlations functions. This is peculiar to the Dyson-Schwinger scheme. We will prove this harmless in the following sections.

We can obtain a reduced set of such equations by using the selected solutions \cite{Frasca:2019ysi} 
\begin{equation}
G_{1\nu}^a(x)\rightarrow\eta_\nu^a\phi(x)
\end{equation}
being $\phi(x)$ a scalar field, and we introduce the $\eta$-symbols with the following properties
\begin{eqnarray}
\eta_\mu^a\eta^{a\mu} &=& N^2-1. \nonumber \\ 
\eta_\mu^a\eta^{b\mu} &=& \delta_{ab}, \nonumber \\
\eta_\mu^a\eta_\nu^a &=& \left(g_{\mu\nu}-\delta_{\mu\nu}\right)/2.
\end{eqnarray}
This yields the set of reduced 1P-function equations
\begin{eqnarray}
&&\partial^2\phi(x)+2Ng^2\Delta(0)\phi(x)+Ng^2\phi^3(x) 
\nonumber \\
&&=\frac{1}{N^2-1}\left[g\sum_{q,i}\eta^{a\nu}\gamma_\nu T^aS_{q}^{ii}(0)\right. \nonumber \\
&&\left.+g\sum_{q,i}{\bar q}_1^i(x)\eta^{a\nu}\gamma_\nu T^a q_1^i(x)\right]
\nonumber \\
&&(i\slashed\partial-{\hat M}_q^i)q_{1}^{i}(x)+g{\bm T}\cdot\slashed\eta\phi(x) q_{1}^{i}(x) = 0.
\end{eqnarray}

We now give here equations for the 2P-functions. We make the choice for the gluon 2P-function
\begin{equation}
    G_{2\mu\nu}^{ab}(x-y)=\left(\eta_{\mu\nu}-\frac{\partial_\mu\partial_\nu}{\partial^2}\right)\Delta_\phi(x-y)
\end{equation}
where $\eta_{\mu\nu}$ is the Minkowski metric, and $\Delta_\phi(x-y)$ is the propagator of the $\phi$ we introduced above to solve the equations to map them onto. Then, the set of remapped 2P-functions is
\begin{eqnarray}
&&\partial^2\Delta_\phi(x-y)+2Ng^2\Delta_\phi(0)\Delta_\phi(x-y)+3Ng^2\phi^2(x)\Delta_\phi(x-y) \nonumber \\
&&=g\sum_{q,i}{\bar Q}^{ia}_\nu(x-y)\gamma^\nu T^a q_{1}^{i}(x)
\nonumber \\
&&+g\sum_{q,i}{\bar q}_1^{i}(x)\gamma^\nu T^a Q^{ia}_\nu(x-y) + \delta^4(x-y)\nonumber \\
&&\partial^2 P^{ad}_2(x-y)=\delta_{ad}\delta^4(x-y) \nonumber \\
&&(i\slashed\partial-{\hat M}_q^i)S^{ij}_q(x-y) \nonumber \\
&&+g{\bm T}\cdot\slashed\eta\phi(x) S^{ij}_q(x-y)=\delta_{ij}\delta^4(x-y) \nonumber \\  
&&\partial^2W_{q\nu}^{ai}(x-y)+2Ng^2\Delta_\phi(0)W_{q\nu}^{ai}(x-y)+3Ng^2\phi^2(x)W_{q\nu}^{ai} \nonumber \\
&&=g\sum_{j}{\bar q}_1^{j}(x)\gamma_\nu T^a S^{ji}_q(x-y)\nonumber \\
&&(i\slashed\partial-{\hat M}_q^i)Q^{ia}_\mu(x-y)+g{\bm T}\cdot\slashed\eta\phi(x) Q^{ia}_\mu(x-y) \nonumber \\
&&+gT^a\gamma_\mu\Delta_\phi(x-y) q_{1}^{i}(x)=0.
\end{eqnarray}

\section{'t Hooft limit}

't Hooft limit corresponds to solve the theory when \cite{tHooft:1973alw, tHooft:1974pnl}
\begin{equation}
  N\rightarrow\infty,\qquad Ng^2=constant, \qquad Ng^2\gg 1.
\end{equation}
We are assuming a SU(N) gauge group and so, $N$ is the number of colors. Therefore, we are able to evaluate our set of Dyson-Schwinger equations in this limit. We need a perturbation series for a coupling formally running to infinity as the one proposed in Ref.\cite{Frasca:2013tma}. To this aim, 
we re-scale $x\rightarrow\sqrt{Ng^2}x$. So, e.g., the equation for the gluon field will become
\begin{eqnarray}
      \partial^2\phi(x')+2\Delta_\phi(0)\phi(x')+3\phi^3(x')&=& \\
\frac{1}{\sqrt{Ng^2}\sqrt{N}(N^2-1)}\left[\sum_{q,i}\eta\cdot\gamma\cdot TS_{q}^{ii}(0)+\right.&& \nonumber \\
\left.\sum_{q,i}{\bar q}_1^i(x')\eta\cdot\gamma\cdot T q_1^i(x')\right].&&\nonumber
\end{eqnarray}
Then, taking formally the 't Hooft limit, it yields the 1P-equations at the leading order
\begin{eqnarray}
	\partial^2\phi_0(x)+2Ng^2\Delta_\phi(0)\phi_0(x)+3Ng^2\phi_0^3(x)=0,& \nonumber \\
		(i\slashed\partial-{\hat M}_q^i){\hat q}_{1}^{i}(x)=0.&
\end{eqnarray}
From this equations we see that the effect of the interactions is on the masses. We can solve this set of equations as
\begin{eqnarray}
\phi_0(x)=\sqrt{\frac{2\mu^4}{m^2+\sqrt{m^4+2Ng^2\mu^4}}}\times
\nonumber \\
{\rm sn}\left(p\cdot x+\chi,\kappa\right),
\end{eqnarray}
being sn a Jacobi elliptical function, $\mu$ and $\chi$ arbitrary integration constants and $m^2=2Ng^2\Delta_\phi(0)$. Then,
\begin{equation}
\kappa=\frac{-m^2+\sqrt{m^4+2Ng^2\mu^4}}{-m^2-\sqrt{m^4+2Ng^2\mu^4}}.
\end{equation}
This is true provided that the following dispersion relation holds
\begin{equation}
    p^2=m^2+\frac{Ng^2\mu^4}{m^2+\sqrt{m^4+2Ng^2\mu^4}}.
\end{equation}
In the same limit we get the set of 2P-equations
\begin{eqnarray}
\partial^2\Delta_\phi(x,y)+2Ng^2\Delta_\phi(0)\Delta(x-y)+3Ng^2\phi_0^2(x)\Delta_\phi(x-y) \nonumber \\
=g\sum_{q,i}{\bar Q}^{ia}_\nu(x,y)\gamma^\nu T^a {\hat q}_{1}^{i}(x)
\nonumber \\
+g\sum_{q,i}{\bar{\hat q}}_1^{i}(x)\gamma^\nu T^a Q^{ia}_\nu(x,y)
+ \delta^4(x-y) \nonumber \\
\partial^2 P^{ad}_2(x-y)=\delta_{ad}\delta^4(x-y) \nonumber \\
(i\slashed\partial-{\hat M}_q^i){\hat S}^{ij}_q(x-y)=\delta_{ij}\delta^4(x-y) \nonumber \\  
\partial^2W_{q\nu}^{ai}(x,y)+2Ng^2\Delta_\phi(0)W_{q\nu}^{ai}(x,y)+3Ng^2\phi_0^2(x)W_{q\nu}^{ai}(x,y)\nonumber \\
=g\sum_{j}{\bar {\hat q}}_1^{j}(x)\gamma_\nu T^a {\hat S}^{ji}(x-y) \nonumber \\
(i\slashed\partial-{\hat M}_q^i){\hat Q}^{ia}_\mu(x,y)+gT^a\gamma_\mu\Delta_\phi(x-y) {\hat q}_{1}^{i}(x)=0.
\end{eqnarray}
In order to solve this set of equations, we consider
\begin{eqnarray}
\partial^2\Delta_0(x-y)+[m^2+3Ng^2\phi_0^2(x)]\Delta_0(x-y)&=&
\nonumber \\
\delta^4(x-y)&&
\end{eqnarray}
that admits the following solution in momenta space \cite{Frasca:2015yva,Frasca:2013tma}
\begin{eqnarray}
   \Delta_0(p)=M{\hat Z}(\mu,m,Ng^2)\frac{2\pi^3}{K^3(\kappa)}\times \nonumber \\
	\sum_{n=0}^\infty(-1)^n\frac{e^{-(n+\frac{1}{2})\pi\frac{K'(\kappa)}{K(\kappa)}}}
	{1-e^{-(2n+1)\frac{K'(\kappa)}{K(\kappa)}\pi}}\times \nonumber \\
	(2n+1)^2\frac{1}{p^2-m_n^2+i\epsilon}
\end{eqnarray}
being
\begin{equation}
M=\sqrt{m^2+\frac{Ng^2\mu^4}{m^2+\sqrt{m^4+2Ng^2\mu^4}}},
\end{equation}
and ${\hat Z}(\mu,m,Ng^2)$ a given constant. This gives rise to a gap equation for the mass shift $m$ on the theory spectrum $m_n$ \cite{Frasca:2017slg}. Given the gluon propagator, one gets
\begin{eqnarray}
{\hat S}^{ij}_q(x,y)=\delta_{ij}(i\slashed\partial-{\hat M}_q^i)^{-1}\delta^4(x-y)\nonumber \\
{\hat Q}^{ia}_\mu(x,y)=-g\int d^4y'\sum_j{\hat S}^{ij}_q(x-y')T^a\gamma_\mu\Delta_0(y',y) {\hat q}_{1}^{j}(y') \nonumber \\
W_{q\nu}^{ai}(x,y)=g\int d^4y'\Delta_0(x-y')\sum_{j}{\bar {\hat q}}_1^{j}(y')\gamma_\nu T^a {\hat S}^{ji}_q(y'-y) \nonumber \\
\Delta(x,y)=\Delta_0(x-y)+ \nonumber \\
g\int d^4y'\Delta_0(x-y')\left[\sum_{q,i}{\bar {\hat Q}}^{ia}_\nu(y',y)\gamma^\nu T^a {\hat q}_{1}^{i}(y')\right. \nonumber \\
\left.+\sum_{q,i}{\bar{\hat q}}_1^{i}(y')\gamma^\nu T^a {\hat Q}^{ia}_\nu(y',y)\right].
\end{eqnarray}

Finally, the quark propagator can be obtained by this set of equations as
\begin{eqnarray}
(i\slashed\partial-{\hat M}_q^i){\hat q}_{1}^{i}(x)= 0\nonumber \\
(i\slashed\partial-{\hat M}_q^i){\hat S}^{ij}_q(x-y)=\delta_{ij}\delta^4(x-y),
\end{eqnarray}
given the mass matrix
\begin{eqnarray}
\label{eq:se}
{\hat M}_q^i=m_qI-
g^2\int d^4y'\Delta_0(x-y')T^a\gamma^\nu\times \nonumber \\
\sum_{k}{\bar {\hat q}}_1^{k}(y')\gamma_\nu T^a {\hat S}^{ki}_q(y'-x).
\end{eqnarray}
This can be solved by iteration starting from the free quark propagator. When the on-shell condition fails, we will have quark confinement.

\section{Non-local NJL approximation}

From eq.(\ref{eq:se}) we can define the the self-energy
\begin{eqnarray}
\Sigma(x,x)=g^2\int d^4y'\Delta_0(x-y')T^a\gamma^\nu\times  \nonumber \\
\sum_{k}{\bar {\hat q}}_1^{k}(y')\gamma_\nu T^a {\hat S}^{ki}_q(y'-x)
\end{eqnarray}
From this, we can introduce a non-local-Nambu-Jona-Lasinio model (nlNJL) \cite{Frasca:2019ysi}
\begin{equation}
(i\slashed\partial-m_q+\Sigma^i_{NJL}(x,x)){\hat S}^{ij}_q(x-y)=\delta_{ij}\delta^4(x-y)
\end{equation}
being now the quark self-energy computed at the first iteration by the free quark propagator giving
\begin{eqnarray}
\Sigma^i_{NJL}(x,x)=g^2\int d^4y'\Delta_0(x-y')T^a\gamma^\nu\times
    \nonumber \\
\sum_{j}{\bar {\hat q}}_0^{j}(y')\gamma_\nu T^a {\hat S}^{ji}_{0q}(y'-x)
\end{eqnarray}
or, in momentum space,
\begin{eqnarray}
\Sigma^i_{NJL}(p)=g^2\int\frac{d^4p_1}{(2\pi)^4}\Delta_0(p_1)T^a\gamma^\nu\times  \nonumber \\
\sum_{j}{\bar {\hat q}}_0^{j}(p)\gamma_\nu T^a {\hat S}^{ji}_{0q}(p_1-p).
\end{eqnarray}

\begin{figure}[H]
%\centering
%\includegraphics[width=.49\textwidth]{surf0}\hfil
%\includegraphics[width=.49\textwidth]{curves}
\includegraphics[width=.49\textwidth]{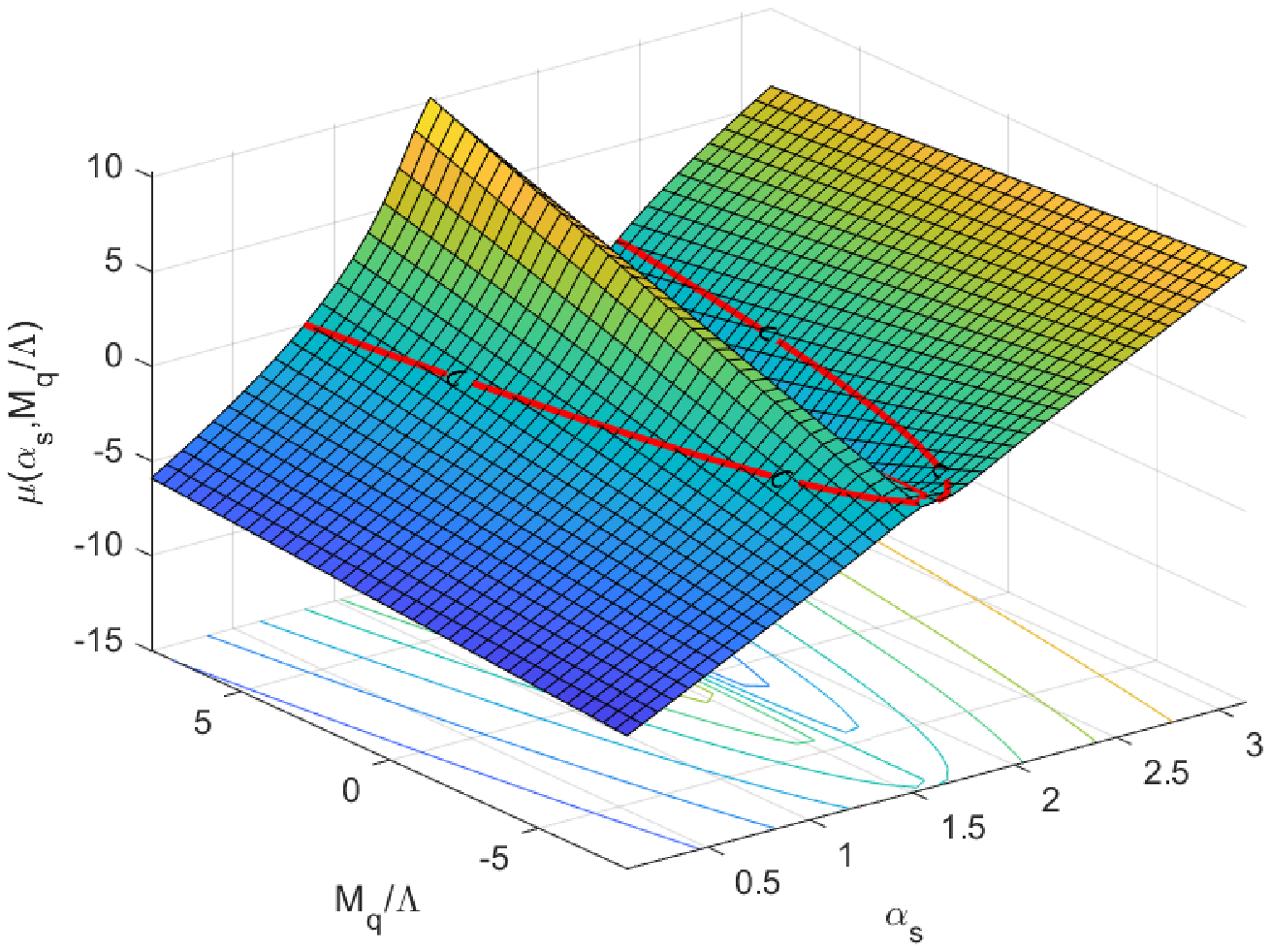}\hfil
\includegraphics[width=.49\textwidth]{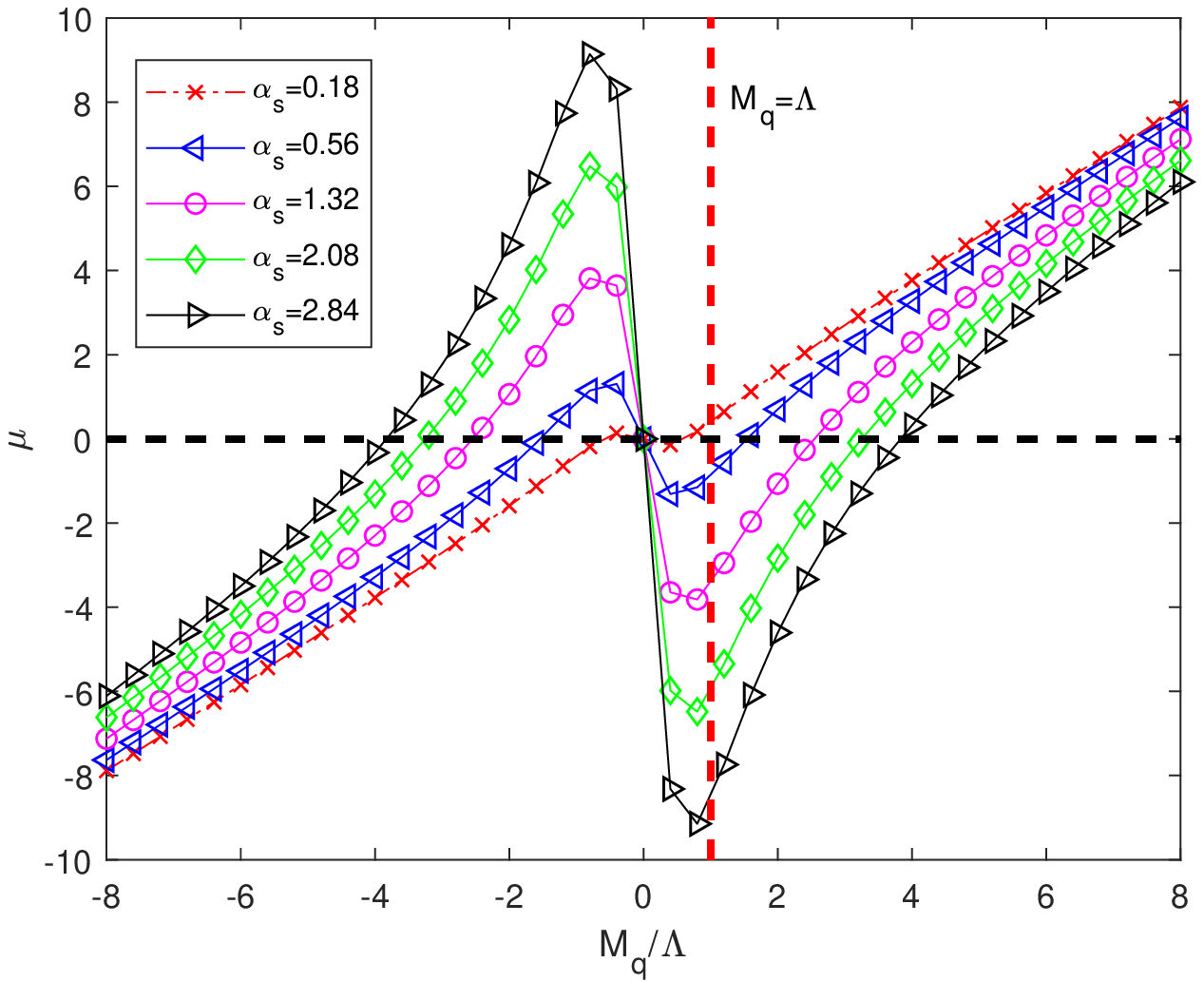}
\caption{Zeros of the quark mass function $\mu$ given by the red curve for the 3D figure (above) and the corresponding profiles with the physical limit in red (below).}\label{fig1}
\end{figure}

This nlNJL-model gives rise, as usual, to a mass gap equation for quarks implying the formation of a condensate. This generally  grants a pole in the quark propagator. Failing to find such a pole means that the quark propagator has no physical mass states and the quarks are confined \cite{Roberts:1994dr,Gribov:1998kb}.
Then, at very low energies
\begin{equation}
M_q=m_q-{\rm Tr}\Sigma^i_{NJL}(0)
\end{equation}
where the trace is over flavors, colors and spinor indexes. This yields
\begin{eqnarray}
M_q=m_q+\frac{N_f(N^2-1)Ng^2}{2}\times \\
\int\frac{d^4p}{(2\pi)^4}\Delta_0(p)\frac{M_q}{p^2+M^2_q}
\nonumber
\end{eqnarray}
We consider the gluon propagator neglecting the mass shift, as this is generally small as shown in \cite{Frasca:2017slg},
\begin{equation}
\Delta_0(p)=\sum_{n=0}^\infty\frac{B_n}{p^2+m_n^2}.
\end{equation}
Therefore, we have to compute
\begin{eqnarray}
M_q=m_q+\frac{N_f(N^2-1)Ng^2}{2}\int\frac{d^4p}{(2\pi)^4}
\times \\
\sum_{n=0}^\infty\frac{B_n}{p^2+m_n^2}\frac{M_q}{p^2+M^2_q}.
\nonumber
\end{eqnarray}
The corresponding integral can be evaluated exactly when a cut-off $\Lambda$ is used, as usual for Nambu-Jona-Lasinio models that are generally expected not to be renormalizable \cite{Klevansky:1992qe}. This yields
\begin{eqnarray}
M_q=m_q+\frac{N_f(N^2-1)Ng^2}{16\pi^2}
\sum_{n=0}^\infty\frac{B_nM_q}{2(m_n^2-M_q^2)}\times
%}} 
\nonumber \\
\left[m_n^2\ln\left(1+\frac{\Lambda^2}{m_n^2}\right)-M_q^2\ln\left(1+\frac{\Lambda^2}{M_q^2}\right)\right].
\end{eqnarray}
This equation is amenable to a numerical treatment provided that $M_q\ge m_q$ and $M_q\ll\Lambda$, $\Lambda$ is the nlNJL-model cut-off. The ultraviolet cut-off represents, at least, the boundary of the region where asymptotic freedom starts to set in (generally taken at $\Lambda\approx 1\ {\rm GeV}$). We normalize the mass function taking $x=m_0/\Lambda$ and $y=M_q/\Lambda$ having set $m_n=(2n+1)m_0$. The mass gap $m_0$ can be assumed to be that of the $\sigma$ meson or f(500) that we fix to $m_0=0.417\ {\rm GeV}$ \cite{Zyla:2020zbs}.
Then, the function to study is
\begin{eqnarray}
\label{eq:y}
y=\frac{m_q}{\Lambda}+\kappa\alpha_s\sum_{n=0}^\infty\frac{B_ny}{(2n+1)^2x^2-y^2}\times \nonumber \\
\left[(2n+1)^2x^2\ln\left(1+\frac{1}{(2n+1)^2x^2}\right)-y^2\ln\left(1+\frac{1}{y^2}\right)\right].
\end{eqnarray}
From this, we derive the mass function
\begin{eqnarray}
\mu(\alpha_s,y)=y-
\frac{m_q}{\Lambda}-\kappa\alpha_s\sum_{n=0}^\infty\frac{B_ny}{(2n+1)^2x^2-y^2}\times \\
\left[(2n+1)^2x^2\ln\left(1+\frac{1}{(2n+1)^2x^2}\right)-y^2
\ln\left(1+\frac{1}{y^2}\right)\right]. \nonumber
\end{eqnarray}
that we plotted in fig.\ref{fig1} for its zeros.

Zero curve exists then, there is a range of physical parameters where the chiral symmetry is broken.
At energies higher than 1 GeV, asymptotic freedom starts to set in and quarks retain their bare masses.
At increasing $\alpha_s$, the effective quark mass starts to be nonphysical overcoming the cut-off. So, we have no solutions and the quark propagator has no physical poles for a free quark. A confinement condition can be straightforwardly obtained by taking $M_q=\Lambda$, the limit of the physical region. This yields
\begin{equation}
\alpha_s=\min_{q=u,d,s}\frac{1-\frac{m_q}{\Lambda}}{\kappa\xi}.
\end{equation}
%\item[\ ]
%%\vspace{0.5cm} 
%\begin{center}
%\fbox{
%\parbox[c]{10.5cm}{
%Support and enlightening discussions with Marco Ruggieri are gratefully acknowledged.
%}}
%\end{center}
being $\kappa=N_fN(N^2-1)/8\pi$, $ \xi=\xi(m_0/\Lambda)$ afunction just depending on the mass gap and the UV cut-off obtainable by eq.(\ref{eq:y}), and $\alpha_s=g^2/4\pi$.

\section{Conclusions}

We have derived the set of Dyson-Schwinger equations, till to 2P-functions, for QCD with the Bender-Milton-Savage technique.
Then, we were able to solve them in the 't Hooft limit. We recognized that the low-energy limit is given by a nonlocal-Nambu-Jona-Lasinio approximation.
Consequently, in the low-energy limit, we were able to show that the model is confining. A confinement condition was also obtained obtained.

\end{document}